\documentclass[12pt]{article}
\usepackage{bm,url,graphicx} 

\usepackage{scicite}

\usepackage{times}

\newenvironment{sciabstract}{%
\begin{quote} \bf}
{\end{quote}}

\newcounter{lastnote}

\title{Community Structure in Time-Dependent, Multiscale, and Multiplex Networks}

\author
{Peter J. Mucha$^{1,2,\ast}$, Thomas Richardson$^{1,3}$, Kevin Macon$^{1}$,\\ Mason A. Porter$^{4,5}$, and Jukka-Pekka Onnela$^{6}$\\
\\
\normalsize{$^{1}$Carolina Center for Interdisciplinary Applied Mathematics, Department of Mathematics,}\\
\normalsize{University of North Carolina, Chapel Hill, NC 27599-3250, USA}\\
\normalsize{$^{2}$Institute for Advanced Materials, Nanoscience and Technology,}\\
\normalsize{University of North Carolina, Chapel Hill, NC 27599, USA}\\
\normalsize{$^{3}$Operations Research, North Carolina State University, Raleigh, NC 27695, USA}\\
\normalsize{$^{4}$Oxford Centre for Industrial and Applied Mathematics,}\\
\normalsize{Mathematical Institute, University of Oxford, Oxford OX1 3LB, UK}\\
\normalsize{$^{5}$CABDyN Complexity Centre, University of Oxford, Oxford OX1 1HP, UK}\\
\normalsize{$^{6}$Department of Health Care Policy, Harvard Medical School, Boston, MA 02115, USA;}\\
\normalsize{Harvard Kennedy School, Harvard University, Cambridge, MA 02138, USA}\\
\\
\normalsize{$^\ast$To whom correspondence should be addressed; E-mail:  mucha@unc.edu.}
}

\date{}

\begin{document}


\baselineskip24pt


\maketitle



\begin{sciabstract}
Network science is an interdisciplinary endeavor, with methods and applications drawn from across the natural, social, and information sciences.  A prominent problem in network science is the algorithmic detection of tightly-connected groups of nodes known as {communities}. We developed a generalized framework of network quality functions that allowed us to study the community structure of arbitrary multislice networks, which are combinations of individual networks coupled through links that connect each node in one network slice to itself in other slices.  This framework allows one to study community structure in a very general setting encompassing networks that evolve over time, have multiple types of links (multiplexity), and have multiple scales.
\end{sciabstract}



\noindent
The study of graphs, or networks, has a long tradition in fields such as sociology and mathematics, and it is now ubiquitous in academic and everyday settings.  An important tool in network analysis is the detection of mesoscopic structures known as communities (or cohesive groups), which are defined intuitively as groups of nodes that are more tightly connected to each other than they are to the rest of the network \cite{structpnas,comnotices,fortunato2009}.
One way to quantify communities is by a quality function that counts intra-community edges compared to what one would expect at random.  Given the network adjacency matrix $\mathbf{A}$, where the component $A_{ij}$ details a direct connection between nodes $i$ and $j$, one can construct a quality function $Q$ \cite{newmodlong,potts}
for the partitioning of nodes into communities as
$Q = \sum_{ij}\left[A_{ij}-P_{ij}\right]\delta(g_i,g_j)$,
where $\delta(g_i,g_j)=1$ if the community assignments $g_i$ and $g_j$ of nodes $i$ and $j$ are the same and $0$ otherwise, and $P_{ij}$ is the expected weight of the edge between $i$ and $j$ under a specified null model.

The choice of null model is a crucial consideration in studying network community structure \cite{comnotices}, ideally respecting the type of network studied. After selecting a null model appropriate to the network and application at hand, one can use a variety of computational heuristics to assign nodes to communities to optimize the quality $Q$ \cite{comnotices,fortunato2009}.
However, such null models have not been available for time-dependent
networks---one has instead had to use {ad hoc} methods
to piece together the structures obtained at different
times \cite{hopcroft2004,bergerwolf2006,palla2007,fenn2009} or abandon quality functions for an alternative such as the Minimum Description Length principle \cite{sun2007}. While tensor
decompositions\cite{seleekolda} have been used to cluster
network data with different types of connections, no quality-function method has been developed for such multiplex networks.

We developed a methodology to remove these limits, generalizing the determination of community structure via quality functions to multislice networks that are defined by coupling multiple adjacency matrices (see Fig.~1). The connections encoded by the network slices are flexible---they can represent variations across time, across different types of connections, or even community detection of the same network at different scales.  However, the usual procedure for establishing a quality function as a direct count of the intra-community edge weight minus that expected at random fails to provide any contribution from these inter-slice couplings.  Because they are specified by common identifications of nodes across slices, inter-slice couplings are either present or absent by definition, so when they do fall inside communities, their contribution in the count of intra-community edges exactly cancels that expected at random.
In contrast, by formulating a null model in terms of stability of communities under Laplacian dynamics, we have derived a principled generalization of community detection to multislice networks, with a single parameter controlling the inter-slice correspondence of communities.

Important to our method is the equivalence between the modularity quality function \cite{structeval} [with a resolution parameter \cite{potts}] and stability of communities under Laplacian dynamics \cite{lambiotte2008}, which we have generalized to recover the null models for bipartite, directed, and signed networks \cite{som}. First, we obtained the resolution-parameter generalization of Barber's null model for bipartite networks \cite{barber2007} by requiring the independent joint probability contribution to stability in \cite{lambiotte2008} to be conditional on the type of connection necessary to step between two nodes.  Second, we recovered the standard null model for directed networks \cite{arenas2007,leicht2008} (again with a resolution parameter) by generalizing the Laplacian dynamics to include motion along different kinds of connections---in this case, both with and against the direction of a link. By this generalization, we similarly recovered a null model for signed networks \cite{gomez2009}.  Third, we interpreted the stability under Laplacian dynamics flexibly to permit different spreading weights on the different types of links, giving multiple resolution parameters to recover a general null model for signed networks \cite{traag2009}.

We applied these generalizations to derive null models for multislice networks that extend the existing quality-function methodology, including an additional parameter $\omega$ to control the coupling between slices. Representing each network slice $s$ by adjacencies
$A_{ijs}$ between nodes $i$ and $j$, with inter-slice couplings $C_{jrs}$ that connect node $j$ in slice $r$ to itself in slice $s$ (see Fig.~1), we have restricted our attention to unipartite, undirected network slices ($A_{ijs}=A_{jis}$) and couplings ($C_{jrs}=C_{jsr}$), but we can incorporate additional structure in the slices and couplings in the same manner as demonstrated for single-slice null models.  Notating the strengths of each node individually in each slice by $k_{js}=\sum_i A_{ijs}$ and across slices by $c_{js} = \sum_r C_{jsr}$, we define the {multislice strength} by $\kappa_{js}=k_{js}+c_{js}$.  The continuous-time Laplacian dynamics given by $\dot{p}_{is} = \sum_{jr} (A_{ijs}\delta_{sr}+\delta_{ij}C_{jsr})p_{jr}/\kappa_{jr} - p_{is}$ respects the intra-slice nature of $A_{ijs}$ and the inter-slice couplings of $C_{jsr}$. Using the steady state probability distribution $p^*_{jr} = \kappa_{jr}/(2\mu)$, where $2\mu=\sum_{jr} \kappa_{jr}$, we obtained the multislice null model in terms of the probability $\rho_{is|jr}$ of sampling node $i$ in slice $s$ conditional on whether the multislice structure allows one to step from $(j,r)$ to $(i,s)$, accounting for intra- and inter-slice steps separately as
\[
	\rho_{is|jr} p^*_{jr} = \left[\frac{k_{is}}{2m_s}\frac{k_{jr}}{\kappa_{jr}}\delta_{sr} + \frac{C_{jsr}}{c_{jr}}\frac{c_{jr}}{\kappa_{jr}}\delta_{ij}\right] \frac{\kappa_{jr}}{2\mu}\,.
\]
The second term in brackets, which describes the conditional probability of motion between two slices, leverages the definition of the $C_{jsr}$ coupling.  That is, the conditional probability of stepping from $(j,r)$ to $(i,s)$ along an inter-slice coupling is non-zero if and only if $i=j$, and it is proportional to the probability $C_{jsr}/\kappa_{jr}$ of selecting the precise inter-slice link that connects to slice $s$.
Subtracting this conditional joint probability from the linear (in time) approximation of the exponential describing the Laplacian dynamics, we obtained a multislice generalization of modularity (see Supporting Online Material for details):
\[
	Q_\mathrm{multislice} = \frac{1}{2\mu}\sum_{ijsr}\left\{\left(A_{ijs}-\gamma_s\frac{k_{is}k_{js}}{2m_s}\right)\delta_{sr} + \delta_{ij}C_{jsr}\right\} \delta(g_{is},g_{jr})\,,
\]
where we have utilized reweighting of the conditional probabilities, which allows one to have a different resolution $\gamma_s$ in each slice. We have absorbed the resolution parameter for the inter-slice couplings into the magnitude of the elements of $C_{jsr}$, which we suppose for simplicity take binary values $\{0,\omega\}$ indicating absence $(0)$ or presence $(\omega)$ of inter-slice links.

Community detection in multislice networks can then proceed using many of the same computational heuristics that are currently available for single-slice networks [though, as with the standard definition of modularity, one must be cautious about the resolution of communities\cite{resolution} and the likelihood of complex quality landscapes that necessitate caution in interpreting results on real networks\cite{good2009}]. We studied examples that have multiple resolutions [Zachary Karate Club\cite{karate}], vary over time [voting similarities in the U.S. Senate\cite{waugh2009}], or are multiplex [the ``Tastes, Ties, and Time" cohort of university students\cite{T3}].  We provide additional details for each example in the Supplementary Online Material.

We performed simultaneous community detection across multiple resolutions (scales) in the well-known Zachary Karate Club network, which encodes the friendships between 34 members of a 1970s university karate club \cite{karate}. Keeping the same unweighted adjacency matrix across slices ($A_{ijs}=A_{ij}$ for all $s$), the resolution associated to each slice is dictated by a specified sequence of $\gamma_s$ parameters, which we chose to be the 16 values $\gamma_s=\{0.25, 0.5, 0.75, \ldots, 4\}$. In Fig.~2, we depict the community assignments obtained for coupling strengths $\omega=\{0, 0.1, 1\}$ between each neighboring pair of the 16 ordered slices. These results simultaneously probe all scales, including the partition of the Karate Club into four communities at the default resolution of modularity \cite{fortunato2009,richardson2009}.  Additionally, we identified nodes that have an especially strong tendency to break off from larger communities (e.g., nodes 24--29 in Fig.~2).

We also considered roll call voting in the United States Senate across time, from the 1st--110th Congresses, covering the years 1789--2008 and including 1884 distinct Senator IDs \cite{voteview}. We defined weighted connections between each pair of Senators by a similarity between their voting, specified independently for each two-year Congress \cite{waugh2009}. We studied the multislice collection of these 110 networks, with each individual Senator coupled to him/herself when appearing in consecutive Congresses. Multislice community detection uncovered interesting details about the
continuity of individual and group voting trends over time that are
simply not captured by the union of the 110 independent partitions of
the separate Congresses. Figure~3 depicts a partition into 9 communities that we obtained using coupling $\omega=0.5$. The Congresses in which three communities appeared simultaneously are each historically significant: The 4th and 5th
Congresses were the first with political parties; the 10th and 11th
Congresses occurred during the political drama of former Vice President Aaron
Burr's indictment for treason; the 14th and 15th Congresses witnessed
the beginning of changing group structures in the
Democratic-Republican party amidst the dying
Federalist party \cite{waugh2009}; the 31st Congress included the Compromise of 1850;
the 37th Congress occurred during the beginning of the American Civil
War; the 73rd and 74th Congresses followed the landslide 1932 election
amidst the Great Depression; and the 85th--88th Congresses brought the
major American civil rights acts, including the Congressional fights
over the Civil Rights Acts of 1957, 1960, and 1964.

Finally, we also applied multislice community detection to a multiplex network of 1640 college students at a northeastern American university \cite{T3}, including symmetrized connections from the first wave of this data representing (1) Facebook friendships, (2) picture friendships, (3) roommates, and (4) student ``housing group'' preferences.  Because the different connection types are categorical, the natural inter-slice couplings connect an individual in a slice to him/herself in each of the other 3 network slices.  This coupling between categorical slices thus differs from that above that connected only neighboring (ordered) slices. Table~\ref{table:T3} indicates the numbers of communities and the percentages of individuals assigned to 1, 2, 3, or 4 communities across the four types of connections for different $\omega$, as a first investigation of the relative redundancy across the connection types.

In summary, our multislice framework makes it possible to study community structure in a much broader class of networks than was previously possible.  Instead of detecting communities in one static network at a time, our formulation generalizing the Laplacian dynamics approach of Ref.~\cite{lambiotte2008} permits the simultaneous quality-function study of community structure across multiple times, multiple resolution parameter values, and multiple types of links.  We used this method to demonstrate insights in real-world networks that would have been difficult or impossible to obtain without the simultaneous consideration of multiple network slices.  Although our examples included only one kind of variation at a time, our framework applies equally well to networks that have multiple such features (e.g., time-dependent multiplex networks), and we expect multislice community detection to become a powerful tool for studying such systems.

\nocite{acks}

\bibliographystyle{Science}
\bibliography{kingcorr}


\clearpage{}

\begin{table}
\centerline{
\begin{tabular}{c||c|c|c|c|c||}
& & \multicolumn{4}{c||}{Comms per Individual (\%)}\\
$\omega$ & \#Comms & 1 & 2 & 3 & 4\\ \hline
0 & 1036 & 0 & 0 & 0 & 100\\
0.1 & 122 & 14.0 & 40.5 & 37.3 & 8.2\\
0.2 & 66 & 19.9 & 49.1 & 25.3 & 5.7\\
0.3 & 49 & 26.2 & 48.3 & 21.6 & 3.9\\
0.4 & 36 & 31.8 & 47.0 & 18.4 & 2.8\\
0.5 & 31 & 39.3 & 42.4 & 16.8 & 1.5\\
1 & 16 & 100 & 0 & 0 & 0\\ \hline
\end{tabular}}
\caption{Communities in the first wave of the multiplex ``Tastes, Ties, and Time'' network \cite{T3}, using the default spatial resolution ($\gamma=1$) in each of the four slices of data (Facebook friendships, picture friendships, roommates, and housing groups) under various couplings $\omega$ across slices, which changed the number of communities and percentages of individuals assigned on a per slice basis to 1, 2, 3, or 4 communities.}
\label{table:T3}
\end{table}


\clearpage{}

\noindent
Fig.~1: Schematic of a multislice network. Four slices $s=\{1,2,3,4\}$ represented by adjacencies $A_{ijs}$ encode \emph{intra-slice} connections (solid). \emph{Inter-slice} connections (dashed) are encoded by $C_{jrs}$, specifying coupling of node $j$ to itself between slices $r$ and $s$. For clarity, inter-slice couplings are shown for only two nodes and depict two different types of couplings: (1) coupling between neighboring slices, appropriate for ordered slices; and (2) all-to-all inter-slice coupling, appropriate for categorical slices.

\bigskip
\hrule
\bigskip

\noindent
Fig.~2: Multislice community detection of the Zachary Karate Club network\cite{karate} across multiple resolutions. Colors depict community assignments of the 34 nodes (renumbered vertically to group similarly-assigned nodes) in each of the 16 slices (with resolution parameters $\gamma_s=\{0.25, 0.5, \ldots, 4\}$), for $\omega=0$ (top), $\omega=0.1$ (middle), and $\omega=1$ (bottom). Dashed lines bound the communities obtained using Newman-Girvan modularity \cite{structeval}.

\bigskip
\hrule
\bigskip

\noindent
Fig.~3: Multislice community detection of U.S. Senate roll call vote similarities \cite{waugh2009} with $\omega=0.5$ coupling of 110 slices across time (110 two-year Congresses, covering 1789--2008). (A) Colors indicate assignments to 9 communities of the 1884 unique Senators (sorted vertically and connected across Congresses by dashed lines) in each Congress they appear. The dark blue and red communities correspond closely to the modern Democratic and Republican parties, respectively.  Horizontal bars indicate the historical period of each community, with accompanying text enumerating nominal party affiliations of the single-slice nodes (each representing a Senator in a Congress): Pro-Administration (PA), Anti-Administration (AA), Federalist (F), Democratic-Republican (DR), Whig (W), Anti-Jackson (AJ), Adams (A), Jackson (J), Democratic (D), and Republican (R). Vertical gray bars indicate Congresses in which three communities appeared simultaneously. (B) The same assignments according to state affiliations.

\clearpage{}
\thispagestyle{empty}

\begin{figure}
\caption{\label{fig:schem}}
\end{figure}

\vspace*{1.0in}

\begin{center}
\includegraphics[width=5.5cm]{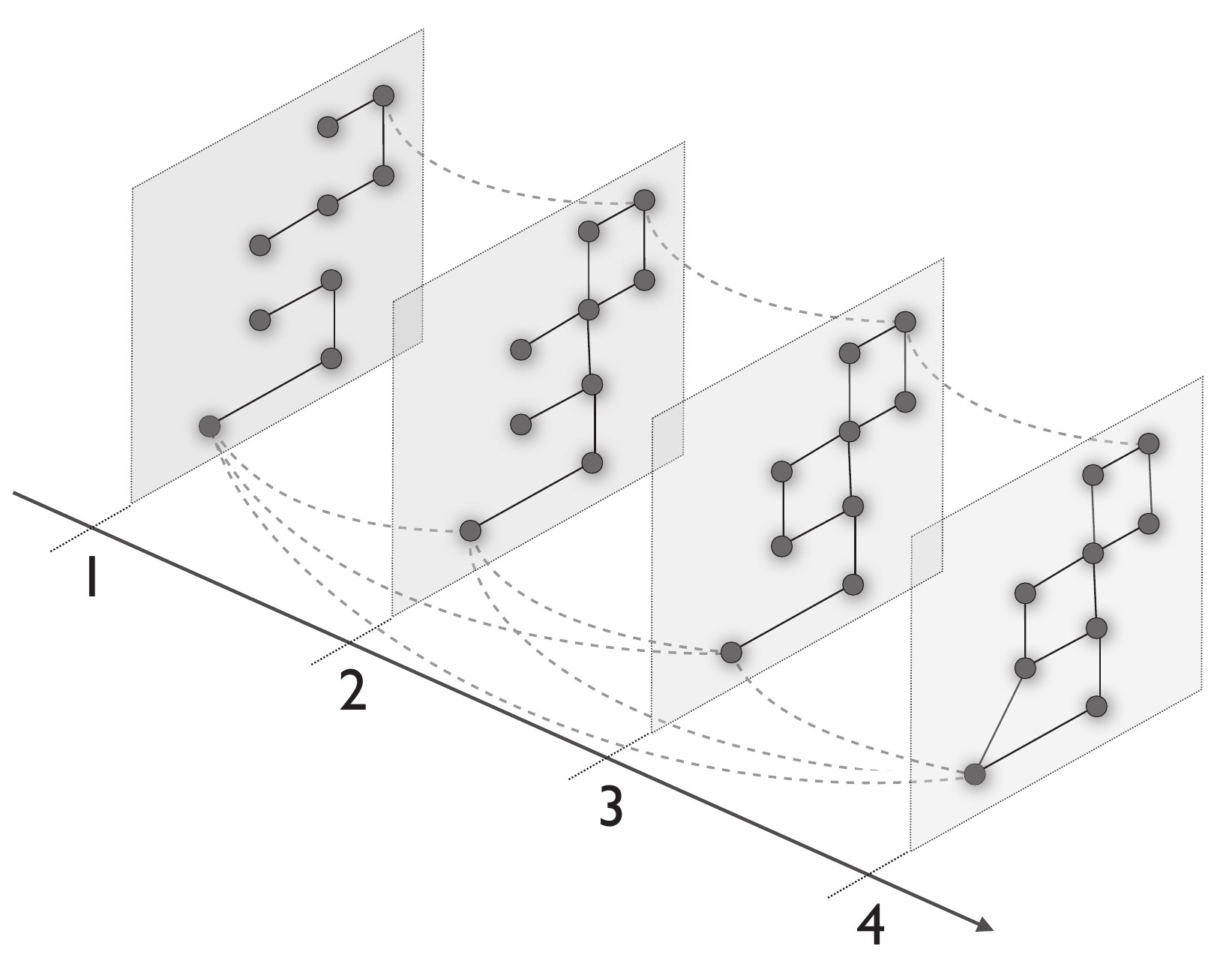}
\end{center}

\clearpage{}
\thispagestyle{empty}

\begin{figure}
\caption{\label{fig:karate}}
\end{figure}

\vspace*{0.1in}

\begin{center}
\includegraphics[width=5.5cm]{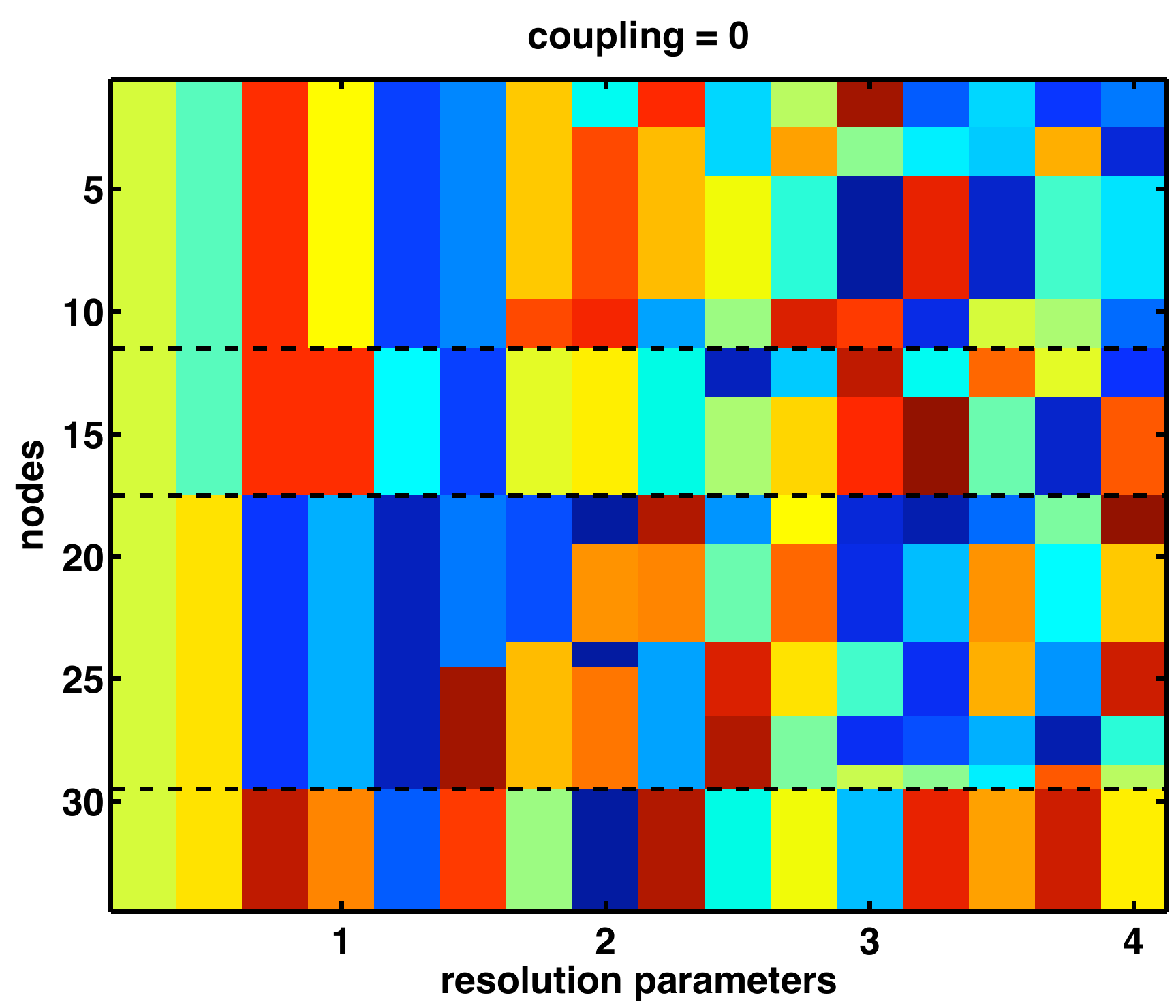}\\
\includegraphics[width=5.5cm]{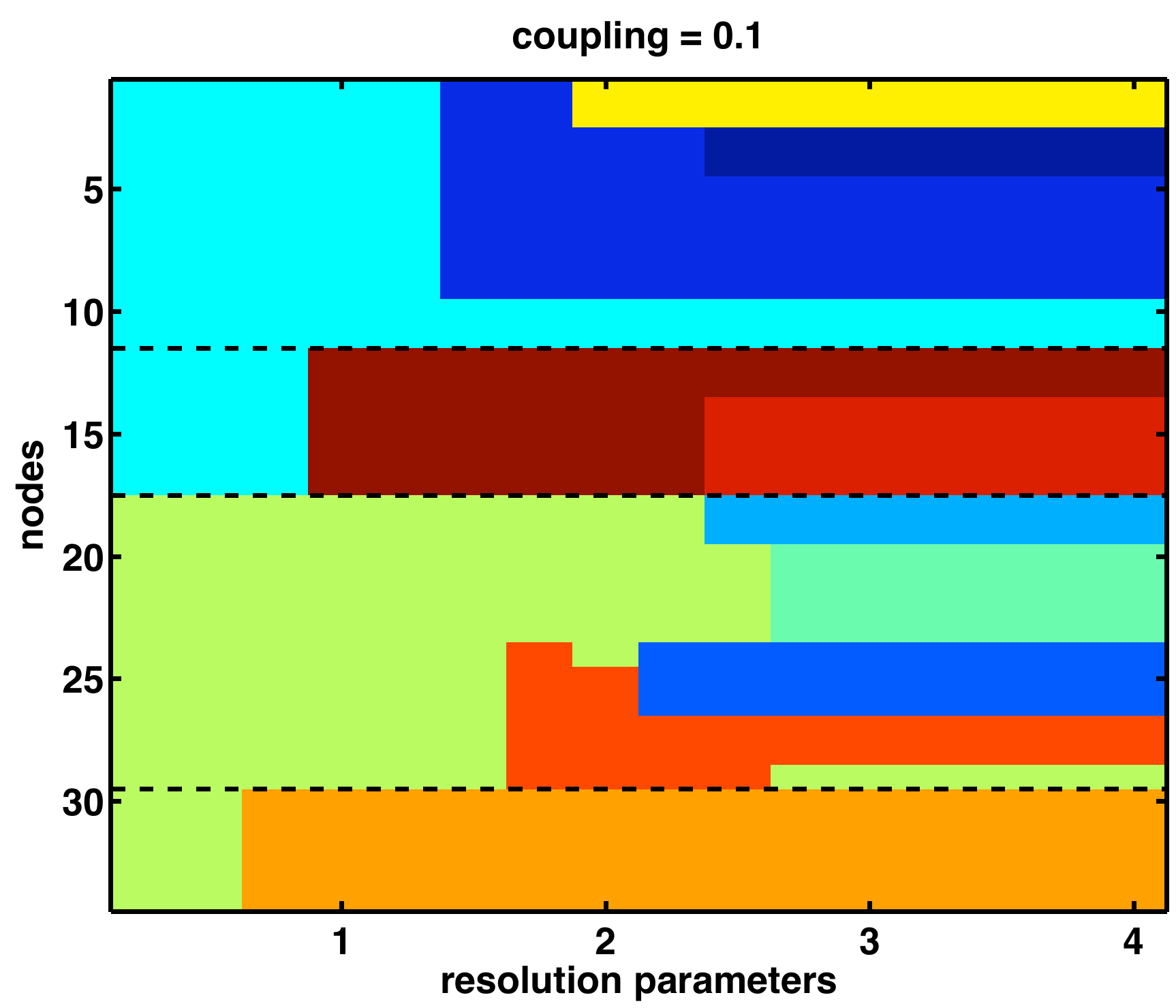}\\
\includegraphics[width=5.5cm]{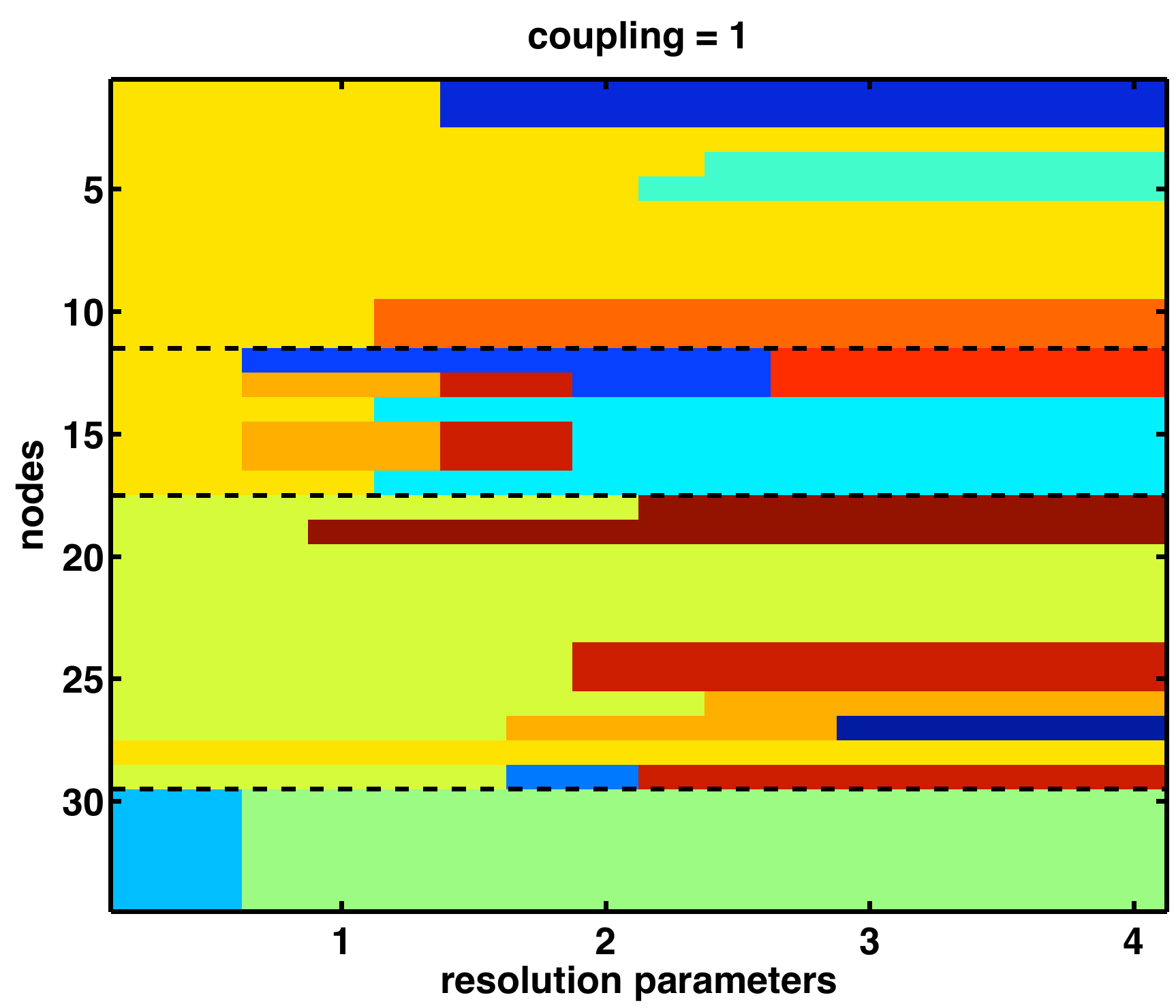}
\end{center}

\clearpage{}
\thispagestyle{empty}

\begin{figure}
\caption{\label{fig:senate}}
\end{figure}

\vspace*{0.1in}

\hspace*{0.875in}{{\bf (A)}}
\vspace*{-0.35in}
\begin{center}
\hspace*{0.12in}\includegraphics[width=9.48cm]{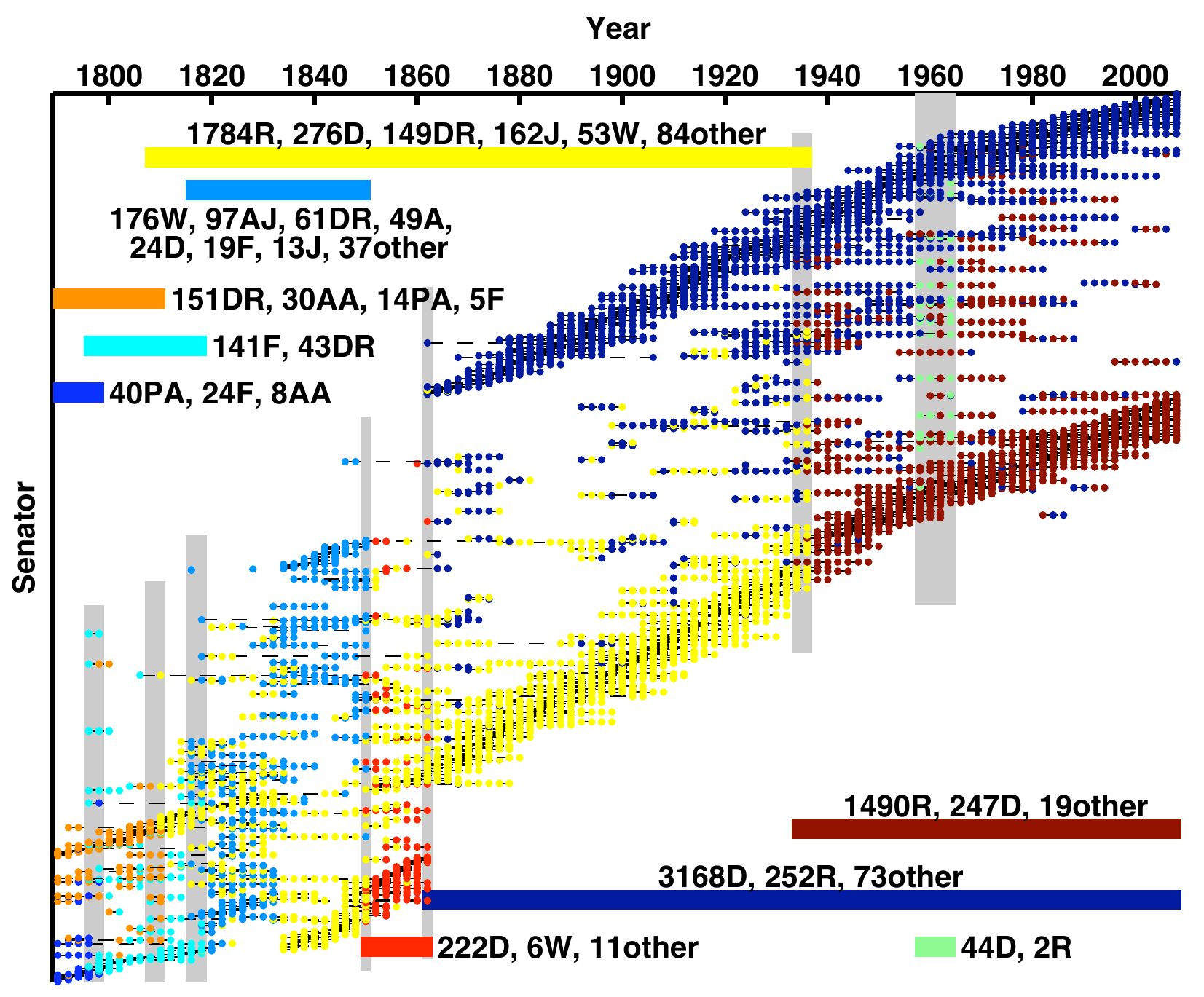}
\end{center}

\vspace*{-0.35in}
\hspace*{0.875in}{{\bf (B)}}
\vspace*{-0.16in}
\begin{center}
\includegraphics[width=10cm]{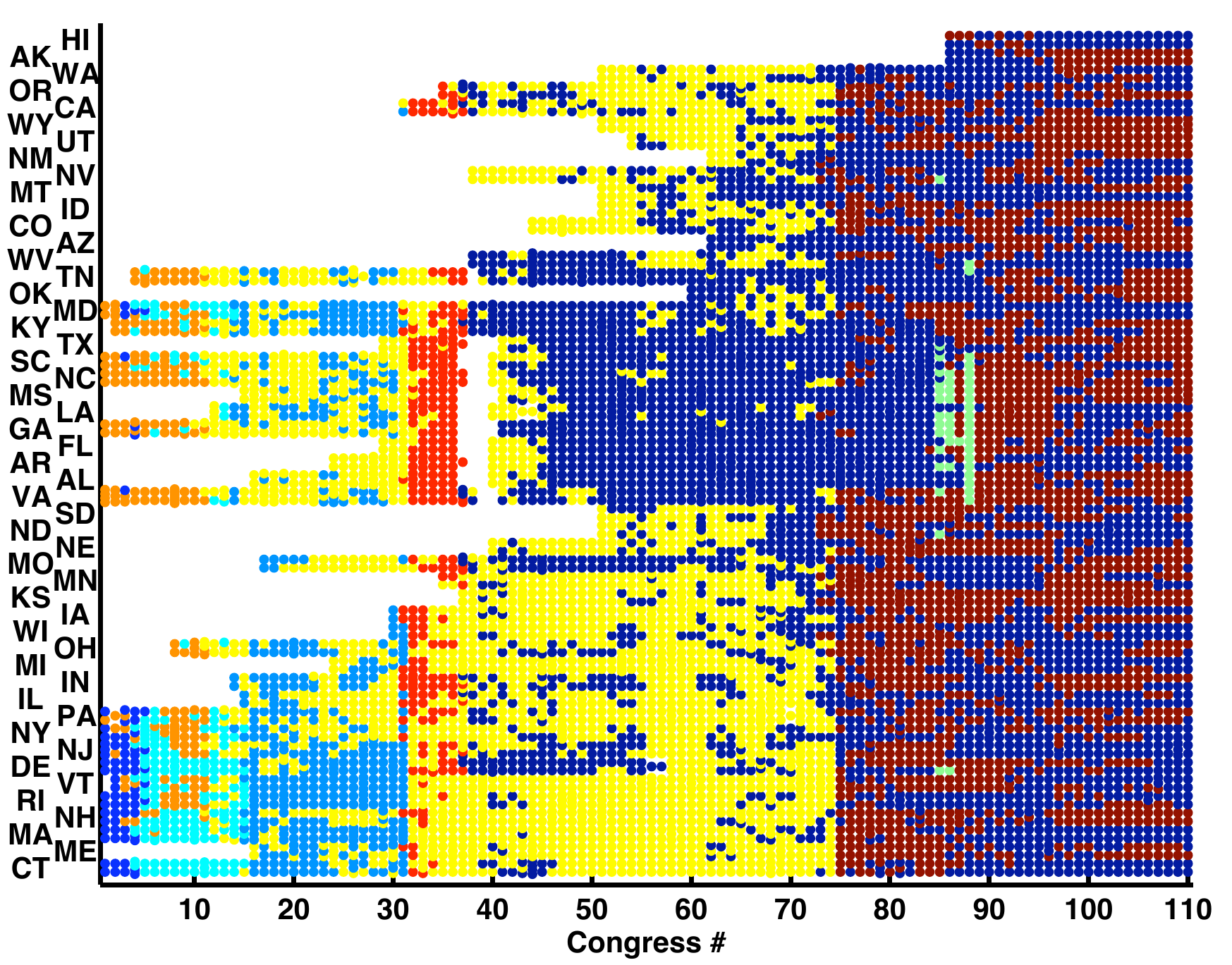}
\end{center}

\clearpage{}
\thispagestyle{empty}

\section*{Supporting Online Material for \emph{Community Structure in Time-Dependent, Multiscale, and Multiplex Networks}}

\author{Peter J. Mucha$^{1,2,\ast}$, Thomas Richardson$^{1,3}$, Kevin Macon$^{1}$,\\ Mason A. Porter$^{4,5}$, and Jukka-Pekka Onnela$^{6}$\\
\\
\normalsize{$^{1}$Carolina Center for Interdisciplinary Applied Mathematics, Department of Mathematics,}\\
\normalsize{University of North Carolina, Chapel Hill, NC 27599-3250, USA}\\
\normalsize{$^{2}$Institute for Advanced Materials, Nanoscience and Technology,}\\
\normalsize{University of North Carolina, Chapel Hill, NC 27599, USA}\\
\normalsize{$^{3}$Operations Research, North Carolina State University, Raleigh, NC 27695, USA}\\
\normalsize{$^{4}$Oxford Centre for Industrial and Applied Mathematics,}\\
\normalsize{Mathematical Institute, University of Oxford, Oxford OX1 3LB, UK}\\
\normalsize{$^{5}$CABDyN Complexity Centre, University of Oxford, Oxford OX1 1HP, UK}\\
\normalsize{$^{6}$Department of Health Care Policy, Harvard Medical School, Boston, MA 02115, USA;}\\
\normalsize{Harvard Kennedy School, Harvard University, Cambridge, MA 02138, USA}\\
\\
\normalsize{$^\ast$To whom correspondence should be addressed; E-mail:  mucha@unc.edu.}
}

\clearpage{}
\setcounter{page}{1}

We provide additional details here about the results and examples that we discussed in the main text.  We begin by reviewing salient results from Ref.~\cite{lambiotte2008} concerning the connection between normalized Laplacian dynamics on networks and the modularity quality function for network community structure. We generalized this methodology to reproduce the null models for bipartite, directed, and signed networks, culminating in the specification of the corresponding quality function for multislice networks, which are combinations of individual networks coupled through links that connect each node in one network slice to itself in other slices. The stacking of multiple slices, linked together by identity arcs, provides a useful representation for visualization and extension of network measures to dynamic graphs (\emph{S1}). We developed a methodology for community detection in such multislice networks, derived from stability of communities under normalized Laplacian dynamics. We additionally considered a similar analysis following from the standard (unnormalized) Laplacian dynamics results from Ref.~\cite{lambiotte2008}. We proved that the domains of optimization of each network partition are convex in the space of parameters for quality functions that are linear in those parameters and comment on possible consequences of this result.


\subsection*{Laplacian Dynamics Formalism} We review the Laplacian dynamics formalism recently developed by Lambiotte \emph{et al.} \cite{lambiotte2008}.

The crucial insight of Ref.~\cite{lambiotte2008} was to rederive network modularity from the continuous-time normalized Laplacian dynamics $\dot{p}_i = \sum_j\frac{1}{k_j}A_{ij}p_j - p_i$ on a unipartite, undirected network defined by the adjacency matrix components $A_{ij}$ with node strengths $k_i=\sum_j A_{ij}$.  They also introduced a notion of {stability} of communities under such dynamics (\emph{13,S2}) by directly comparing the joint probability at stationarity of independent appearances at nodes $i$ and $j$ with the linear (in time) approximate map from node $j$ to node $i$.  In so doing, they derived a quality function equivalent to Newman-Girvan (NG) modularity \cite{structeval} at unit time and the standard Potts generalization of NG modularity\cite{potts} that includes a resolution parameter (which is then interpreted as an inverse time).

The normalized Laplacian dynamics, $\dot{p}_i = \sum_j\frac{1}{k_j}A_{ij}p_j - p_i$, have a steady state given by $p^*_j = k_j/(2m)$, where $2m=\sum_i k_i=\sum_{ij} A_{ij}$ describes the total strength (i.e., total edge weight) in the network. In Ref.~\cite{lambiotte2008}, Lambiotte \emph{et al.}\ quantified a measure of the {stability} $R(t)$ of a specified partition of the network into communities using the probability that a random walker remains within the same community after time $t$, in statistically steady conditions, relative to that expected under independence. Using the operator $L_{ij} = A_{ij}/k_j - \delta_{ij}$ of the dynamics, where $\delta_{ij}$ is the Kronecker delta, they specified this stability as
\begin{equation}
	R(t) = \sum_{ij}\left[\left(e^{t\mathbf{L}}\right)_{ij} p^*_j - p^*_i p^*_j\right]\delta(g_i,g_j)\,,
	\label{eq:stability}
\end{equation}
where the contribution from an independence assumption appears in the second term in brackets.  Expanding the matrix exponential in equation (\ref{eq:stability}) to first-order in $t$, so that $(e^{t\mathbf{L}})_{ij}\approx \delta_{ij}+tL_{ij}$, Lambiotte \emph{et al.} demonstrated that $R(t)$ directly yields the quality function\cite{lambiotte2008}
\begin{equation}
	Q(t) = \frac{1}{2m}\sum_{ij}\left[tA_{ij} - \frac{k_ik_j}{2m}\right]\delta(g_i,g_j)
\end{equation}
up to $\delta_{ij}$ factors that always contribute to the sum and are thus immaterial in identifying partitions that optimize $Q(t)$.  The resulting quality function reduces to NG modularity for $t=1$. Moreover, they showed that dividing by $t$ (which has no effect on the optima for specified $t$) provides a direct interpretation of the resolution parameter $\gamma=1/t$ when the quality is written in the usual form \cite{potts}: $Q = \frac{1}{2m}\sum_{ij}\left[A_{ij} - \gamma\frac{k_ik_j}{2m}\right]\delta(g_i,g_j)$. Hence, the stability of the community partition relative to that expected under independence provides a natural definition for the null model employed in the quality function.


\subsection*{Generalized Laplacian Dynamics} We extended the formalism of Lambiotte \emph{et al.} \cite{lambiotte2008} to multislice networks by considering three crucial generalizations.

First, we restricted the expected independent contribution given by the probability of a random walker remaining within the same community after time $t$ in the statistically steady state to one {that is conditional on the type of connection necessary to step between two nodes}. That is, we replaced the $p_i^* p_j^*$ independent contribution in equation (\ref{eq:stability}) with a conditional independent contribution $\rho_{i|j} p_j^*$, where $\rho_{i|j}$ is the conditional probability at stationarity of jumping to node $i$ from node $j$ along a specific edge type that is allowed by the specified category of networks.  This constraint on the independent contribution is consistent with the linear-in-time expansion of the exponential map employed in the calculation of the expected joint population, giving $\hat{R}(t) = \sum_{ij}\left[\left(\delta_{ij}+tL_{ij}\right) p^*_j - \rho_{i|j} p^*_j\right]\delta(g_i,g_j)$ after linearization of the exponential. For instance, we considered undirected bipartite networks.  Such networks have two types of nodes (e.g., a person might belong to an organization), and every edge must connect a node of one type to a node of the other.  The adjacency matrix ${\bf A}$ in the operator $L_{ij} = A_{ij}/k_j - \delta_{ij}$ takes a bipartite form (i.e., it consists of off-diagonal blocks), and gives the same formula for the steady state, $p^*_j = k_j/(2m)$.  However, $\rho_{i|j} = b_{ij} k_i/m$, where $b_{ij}$ is an indicator of bipartiteness that is equal to $1$ if nodes $i$ and $j$ are of different types and $0$ otherwise. The denominator in $\rho_{i|j}$ is $m$ (cf.\ $2m$) because the probability of stepping to node $i$ conditional on the additional information that the jump is along an edge going towards a node of $i$'s type doubles the probability. Again neglecting $\delta_{ij}$ contributions, dividing by $t$, and setting $\gamma = 1/t$, we thus obtained
\begin{equation}
	Q_\mathrm{bipartite} = \frac{1}{2m}\sum_{ij}\left[A_{ij} - \gamma b_{ij}\frac{k_ik_j}{m}\right]\delta(g_i,g_j)\,,
\end{equation}
which is the generalization of the ($\gamma=1$) Barber bipartite null model\cite{barber2007} obtained by incorporating the resolution parameter $\gamma$.

Second, we generalized the Laplacian dynamics to include motion along multiple types of connections. For example, we considered a directed network (so that ${\bf A}$ is no longer symmetric) with $k^\mathrm{in}_i = \sum_j A_{ij}$ and $k^\mathrm{out}_j = \sum_i A_{ij}$.  We defined the normalized Laplacian dynamics to include motion equally along both incoming and outgoing edges, subject to the normalization $k_j = k^\mathrm{in}_j + k^\mathrm{out}_j$.  That is, we studied the dynamics $\dot{p}_i = \sum_j L_{ij}p_j = \sum_j\frac{1}{k_j}(A_{ij}+A_{ji})p_j - p_i$, which again has steady state $p^*_j = k_j/(2m)$, with $2m=\sum_j k_j=2\sum_{ij}A_{ij}$. The change induced by the consideration of the directed network occurs in the conditional probability $\rho_{i|j}$, which must respect the type of edge (incoming versus outgoing) that is used to arrive at node $i$ as well as the fraction of such edges available in the departure from node $j$.  We thus obtained
\begin{equation}
	\rho_{i|j}p^*_j = \left(\frac{k^\mathrm{in}_i}{m}\frac{k^\mathrm{out}_j}{k_j} + \frac{k^\mathrm{out}_i}{m}\frac{k^\mathrm{in}_j}{k_j}\right)\frac{k_j}{2m} = \frac{k^\mathrm{in}_i k^\mathrm{out}_j + k^\mathrm{out}_i k^\mathrm{in}_j}{2m^2}\,,
\end{equation}
where each additive term combines the probability of picking a particular type of edge when departing node $j$ with the probability of arriving at node $i$ given that the motion is on that type of edge. Because of the symmetry in summing over $\{i,j\}$ pairs, we have equivalently rewritten the resulting partition quality as
\begin{equation}
	Q_\mathrm{directed} = \frac{1}{m}\sum_{ij}\left[A_{ij} -
\gamma\frac{k^\mathrm{in}_ik^\mathrm{out}_j}{m}\right]\delta(g_i,g_j)\,,
\end{equation}
which (as with bipartite networks) yielded the natural extension of the corresponding standard ($\gamma=1$) null model for directed networks\cite{arenas2007,leicht2008} by incorporation of the resolution parameter $\gamma$.  This approach contrasts with that of Lambiotte \emph{et al.}, which restricted consideration to motion following the link directions (\emph{13,S3}).

We also studied the Laplacian dynamics given by the operator $L_{ij}=(A^+_{ij}+A^-_{ij})/k_j - \delta_{ij}$ (with $k_j=k^+_j+k^-_j$), which similarly yielded a null model for (undirected) signed networks, in which link weights can be either positive or negative.  Edges can be separated into ones that contribute positively (for which $A^+_{ij} \geq 0$) and those that contribute negatively (for which $A^-_{ij}\geq 0$).  As before, the steady state is given by $p^*_j=k_j/(2m)$, where now $m=m^+ + m^-$. Because of the penalizing contribution desired from the $A^-_{ij}\geq 0$ links, we chose to weight the $A^-$ and $k^-$ contributions negatively when they appeared in the partition stability formula, which is given by equation (\ref{eq:stability}).  Aside from this sign convention, we calculated the conditional probability at stationarity using the same procedure as in the directed case, keeping track of whether the movement from node $j$ to node $i$ is along a positive or negative edge.  This gave
\begin{equation}
	\rho_{i|j}p^*_j = \left(\frac{k^{+}_i}{2m^+}\frac{k^{+}_j}{k_j} - \frac{k^{-}_i}{2m^-}\frac{k^{-}_j}{k_j}\right)\frac{k_j}{2m} = \frac{1}{2m}\left(\frac{k^{+}_i k^{+}_j}{2m^+} - \frac{k^{-}_i k^{-}_j}{2m^-}\right)\,,
\end{equation}
and yielded $Q = \frac{1}{2m}\sum_{ij}\left[A^+_{ij} - A^-_{ij} -
\gamma\left(\frac{k^{+}_ik^{+}_j}{2m^+}-\frac{k^{-}_ik^{-}_j}{2m^-}\right)\right]\delta(g_i,g_j)$.  This quality function reduces at $\gamma = 1$ to one proposed signed null model\cite{gomez2009} and is a special case of a more general signed null model\cite{traag2009} that includes separate resolution parameters ($\gamma^+$ and $\gamma^-$) for the positive and negative contributions.  We reconstructed the latter null model by using our third generalization, which we present next.

Our third generalization was to flexibly interpret the stability under Laplacian dynamics in order to permit different spreading weights on the different types of links. This was not an issue in our consideration of directed networks unless one wants to weight incoming and outgoing edges differently.  On the other hand, it might be desirable for signed networks to consider reweighted conditional probabilities at stationarity using some factor other than the relative strengths of the different edges at node $j$ (even though we only considered a single specification of the underlying Laplacian dynamics).  This generalization gave
\begin{equation}
	Q_\mathrm{signed} = \frac{1}{2m}\sum_{ij}\left[A^+_{ij} - A^-_{ij} -
\left(\gamma^+\frac{k^{+}_ik^{+}_j}{2m^+}-\gamma^-\frac{k^{-}_ik^{-}_j}{2m^-}\right)\right]\delta(g_i,g_j)\,,
\end{equation}
with two resolution parameters ($\gamma^+$ and $\gamma^-$), which is the undirected version of the aforementioned more general null model for signed networks \cite{traag2009}, with the full directed version similarly obtained by combining the above generalizations.

Having shown that our generalizations recovered the appropriate null models for other categories of networks (bipartite, directed, and signed), we applied this methodology to the far more general framework of multislice networks. We supposed that each slice $s$ of a network is represented by adjacencies $A_{ijs}$ between nodes $i$ and $j$ and specified inter-slice couplings $C_{jrs}$ that connect node $j$ in slice $r$ to itself in slice $s$ (see Fig.~1). That is, notationally, we used two indices to specify each node-slice: a single node (e.g., $i$) in an indicated slice (e.g., $s$). For simplicity, we restricted our attention to undirected network slices ($A_{ijs}=A_{jis}$) and undirected couplings ($C_{jrs}=C_{jsr}$), but we can incorporate additional structure in the slices and couplings in the same manner as in the single-slice derivations above.  For convenience, we notated the strengths of each node individually in each slice, so that $k_{js}=\sum_i A_{ijs}$, $c_{js} = \sum_r C_{jsr}$, and we defined the \emph{multislice strength} $\kappa_{js}=k_{js}+c_{js}$. We studied a continuous-time Laplacian process analogous to those above that respects the intra-slice nature of $A_{ijs}$ and the inter-slice couplings of $C_{jsr}$, specified by
$\dot{p}_{is} = \sum_{jr} (A_{ijs}\delta_{sr}+\delta_{ij}C_{jsr})p_{jr}/\kappa_{jr} - p_{is}$, which has steady state probability distribution $p^*_{jr} = \kappa_{jr}/(2\mu)$, where $2\mu=\sum_{jr} \kappa_{jr}$. We then specified the associated multislice null model using the
probability $\rho_{is|jr}$ of sampling node-slice $(i,s)$ conditional on whether the multislice structure allows one to step from node-slice $(j,r)$ to node-slice $(i,s)$, considering intra- and inter-slice steps separately:
\begin{equation}
	\rho_{is|jr} p^*_{jr} = \left[\frac{k_{is}}{2m_s}\frac{k_{jr}}{\kappa_{jr}}\delta_{sr} + \frac{C_{jsr}}{c_{jr}}\frac{c_{jr}}{\kappa_{jr}}\delta_{ij}\right] \frac{\kappa_{jr}}{2\mu}\,.
\end{equation}
The first term in brackets above describes the conditional probability appropriate for motion along intra-slice edges, analogous to those in the generalized derivation of other null models above, including the probability, $k_{jr}/\kappa_{jr}$, of using an intra-slice edge when leaving $(j,r)$ and the resulting restriction to the given slice ($\delta_{sr}$) made explicit.
The second term in brackets, which similarly describes the conditional probability of motion between two slices, leverages the known definition of the $C_{jsr}$ coupling.  That is, the conditional probability of stepping from $(j,r)$ to $(i,s)$ along an inter-slice coupling is non-zero if and only if $i=j$, and it is proportional to the probability $C_{jsr}/\kappa_{jr}$ of selecting the precise inter-slice link that connects to slice $s$ from all edges connected to $(j,r)$.  The inter-slice strengths $c_{jr}$ therefore canceled naturally as part of this calculation. Subtracting this conditional joint probability from the linear (in time) approximation of the exponential describing the Laplacian dynamics on the multislice networks, we then obtained a multislice generalization of modularity:
\begin{equation}
	Q_\mathrm{multislice} = \frac{1}{2\mu}\sum_{ijsr}\left\{\left(A_{ijs}-\gamma_s\frac{k_{is}k_{js}}{2m_s}\right)\delta_{sr} + \delta_{ij}C_{jsr}\right\} \delta(g_{is},g_{jr})\,,
\label{eq:multislice}
\end{equation}
where we have again utilized reweighting of the conditional probabilities, allowing for different resolutions $\gamma_s$ in each slice. We absorbed the corresponding resolution parameter for the inter-slice couplings into the magnitude of the elements of $C_{jsr}$, which we then supposed for simplicity take binary values $\{0,\omega\}$ indicating absence/presence of inter-slice links.

In the absence of such a reweighting in the interpretation of the stability of the partition, with $\gamma_s=\gamma$ for all $s$, the corresponding prefactor on $C_{jsr}$ absorbed above is $(1-\gamma)$. Imposing the choice $\gamma=1$ then recovered the usual interpretation of modularity as a count of the total weight of intra-slice edges minus the weight expected at random, and (as expected) the specified deterministic $C_{jsr}$ contribution dropped out entirely, because such inter-slice links are definitional to the multislice network.  In contrast, by leveraging the notion of stability under Laplacian dynamics, generalized appropriately, we have derived a principled generalization of modularity to multislice networks.

Choosing binary-valued $C_{jsr}=\{0,\omega\}$ requires only a single coupling parameter $\omega$ to control the extent of inter-slice correspondence of communities. When $\omega=0$, there is no benefit from extending communities across slices, so the optimal partition is obtained from independent optimization of the corresponding quality function in each slice. At the other extreme, when $\omega$ becomes sufficiently large, the quality-optimizing partitions force the community assignment of a node to remain the same across all slices in which that node appears, and the multislice quality reduces to a difference between the
adjacency matrix summed over the contributions from the individual slices and the sum over the separate single-slice null models (with selected $\gamma_s$). That is, the null model obtained in the limit of large $\omega$ is not the same as the standard NG null model on the adjacency matrix summed across slices, which only relies on the total summed degrees; rather, the required sum of the single-slice null models respects the degree sequences of these different contributions separately. The generality of this framework also allows one to consider different weights across the $C_{jsr}$ couplings, if deemed appropriate for a particular application.  Additionally, we note that the linearity of equation (\ref{eq:multislice}) with respect to the $\{\gamma_s,\omega\}$ parameters necessitates that the modularity-optimizing domain of a single partition is convex in this parameter space (as derived below, with a brief discussion of consequences).


\subsection*{Unnormalized Multislice Laplacian Dynamics} As discussed by Lambiotte \emph{et al.} \cite{lambiotte2008}, a similar analysis of the stability of communities under standard (i.e., unnormalized) Laplacian dynamics can be used to yield a quality function with a null model corresponding to a uniform random graph \cite{potts}. We generalized this result to the multislice setting using a natural definition of the relevant independent probabilities subject to conditions imposed by the network structure specific to our multislice setting (similar to our derivation for normalized Laplacian dynamics).

We specified the standard Laplacian dynamics on a multislice network defined by $A_{ijs}$ and $C_{jsr}$ by $\dot{p}_{is} = \sum_{jr} (A_{ijs}\delta_{sr}+\delta_{ij}C_{jsr})p_{jr}/\langle\kappa\rangle - p_{is}\kappa_{is}/\langle\kappa\rangle$, where angled brackets denote an average over the entire multislice network and we recall that $\kappa_{js}=k_{js}+c_{js}$ is the multislice strength.  The steady-state probability distribution under these dynamics is constant.  Hence, $p^*_{jr} = 1/N$, where $N$ is the total number of nodes summed across slices in the multislice network.  We then scale the conditional probability $\rho_{is|jr}$ of stepping from node $j$ at slice $r$ to node $i$ at slice $s$ appropriate to the selected standard dynamics, where the rate of leaving node $j$ at slice $r$ is proportional to $\kappa_{jr}$ [cf. the constant rate of leaving $(j,r)$ in the normalized Laplacian dynamics in the rest of this paper].  Given that we repeated the procedure of allowing different resolution parameters (inverse times) both within and across slices, it was sufficient for us to consider the conditional independent probability in the form
\begin{equation}
	\rho_{is|jr} p^*_{jr} = \left[\delta_{sr} + {C_{jsr}}\delta_{ij}\right] \frac{1}{N}\,.
\end{equation}
Ignoring $\delta_{ij}\delta_{sr}$ contributions to quality, which have no effect on identifying the optimal partition, we obtained
\begin{equation}
	Q = \sum_{ijsr}\left\{\left(A_{ijs}-\gamma_s\right)\delta_{sr} + \delta_{ij}C_{jsr}\right\} \delta(g_{is},g_{jr})
\end{equation}
as the multislice generalization of the uniform random null model.  Note that we once again absorbed the inter-slice coupling strength directly into the binary values of $C_{jsr}=\{0,\omega\}$.  (Again, if desired, one can also consider different weights across the $C_{jsr}$ couplings.)  As with the multislice null model that we obtained from normalized Laplacian dynamics, the limiting behaviors of this quality function are towards independent partitioning of each slice as $\omega\to 0$ and towards averaging over slices for $\omega\gg 1$, though the latter is greatly simplified here since it is merely a sum over constant contributions, in contrast with the more detailed null model in the large coupling limit corresponding to normalized Laplacian dynamics.


\subsection*{Convex Domains of Optimization} We proved that the linearity of equation (\ref{eq:multislice}) with respect to the $\{\gamma_s,\omega\}$ parameters necessitates that the quality-optimizing domain of a single partition be convex in this parameter space. This result holds more generally for any community-detection quality function that is linear in its parameters. That is, if an identified partition of the network is the highest-quality partition at two points in parameter space, then it necessarily gives the best partition along the entire line segment connecting those two points.

The proof of this convexity result followed from the consideration of a line in parameter space that contains two distinct optima at different points. For the purposes of this proof, we notated the parameters (e.g., resolution parameters and/or inter-slice coupling strengths) by the vector array $\boldsymbol\lambda$ and the modularity-like quality function as
\begin{equation}
	Q  = \sum_{ij} B_{ij}\delta(g_i,g_j)
= \sum_{ijp}\left[A_{ij}-\lambda_pP_{ijp}\right]\delta(g_i,g_j)\,,
\label{eq:qualitylinear}
\end{equation}
where the notation $i$ and $j$ for the node indices naturally generalizes over the complete multislice network. That is, $Q = \mathbf{B}\boldsymbol{:\chi} = (\mathbf{A}-\boldsymbol{\lambda\cdot}\mathbf{P})\boldsymbol{:\chi}$, where $\boldsymbol\chi$ is the common-community indicator with elements $\delta(g_i,g_j)$ specific to the selected partition.  The meaning of the double contractions (e.g., $\mathbf{B}\boldsymbol{:\chi}$) over indices and dot products over parameters ($\boldsymbol{\lambda\cdot}\mathbf{P}$) is clear from equation (\ref{eq:qualitylinear}).

We then assumed without loss of generality that a partition specified by $\boldsymbol\chi_1$ is the unique optimum for parameters $\boldsymbol\lambda_1$, with $A_1 = \mathbf{A}\boldsymbol{:\chi}_1$ and $\mathbf{P}_1 = \mathbf{P}\boldsymbol{:\chi}_1$ defined so that $Q_1 = A_1 - \boldsymbol\lambda_1\boldsymbol\cdot\mathbf{P}_1$. If the distinct partition specified by $\boldsymbol\chi_2$ is strictly optimal to $\boldsymbol\chi_1$ for parameters $\boldsymbol\lambda_2$ (with analogous definitions for $Q_2 = A_2 - \boldsymbol\lambda_2\boldsymbol\cdot\mathbf{P}_2$), then we showed it must follow that
\begin{equation}
	Q_1 = A_1 - \boldsymbol\lambda_1\boldsymbol\cdot\mathbf{P}_1
> A_2 - \boldsymbol\lambda_1\boldsymbol\cdot\mathbf{P}_2
\quad\mbox{and}\quad
	Q_2 = A_2 - \boldsymbol\lambda_2\boldsymbol\cdot\mathbf{P}_2
> A_1 - \boldsymbol\lambda_2\boldsymbol\cdot\mathbf{P}_1\,.
\end{equation}
We combined these inequalities to yield
$(\boldsymbol\lambda_2-\boldsymbol\lambda_1)\boldsymbol\cdot\mathbf{P}_1 > (\boldsymbol\lambda_2-\boldsymbol\lambda_1)\boldsymbol\cdot\mathbf{P}_2$.  We then considered a vector array $\boldsymbol\lambda_3$ that is colinear with $\boldsymbol\lambda_2$ and $\boldsymbol\lambda_1$, so that $\boldsymbol\lambda_3 = \boldsymbol\lambda_2 + f(\boldsymbol\lambda_2-\boldsymbol\lambda_1)$ with $f>0$, which yielded the result that the quality of the $\boldsymbol\chi_1$ and
$\boldsymbol\chi_2$ partitions at $\boldsymbol\lambda_3$ must satisfy
\begin{equation}
	A_2 - \boldsymbol\lambda_3\boldsymbol\cdot\mathbf{P}_2
= A_2 - \boldsymbol\lambda_2\boldsymbol\cdot\mathbf{P}_2 -
f(\boldsymbol\lambda_2-\boldsymbol\lambda_1)\boldsymbol\cdot\mathbf{P}_2
> A_1 - \boldsymbol\lambda_2\boldsymbol\cdot\mathbf{P}_1 -
f(\boldsymbol\lambda_2-\boldsymbol\lambda_1)\boldsymbol\cdot\mathbf{P}_1
= A_1 - \boldsymbol\lambda_3\boldsymbol\cdot\mathbf{P}_1\,.
\end{equation}
That is, the partition $\boldsymbol\chi_2$ is necessarily of higher quality than $\boldsymbol\chi_1$ at $\boldsymbol\lambda_3$ (though neither of them needs to be the optimum there).  Therefore, non-convex domains of optimization are forbidden in the parameter space of quality functions of the form in equation (\ref{eq:qualitylinear}).

This requirement of convex domains of quality optimization might be useful for comparing results across different resolution and coupling parameters, not only in the present multislice setting but for {any} network-partitioning quality function that is linear in resolution parameters.  Although other quality functions might of course be considered, we note that each quality function discussed in the present manuscript is of the general form in equation (\ref{eq:qualitylinear}).  Computational results that do not conform to convex domains of optimization typically indicate regions in which further computation should uncover better optima.  Indeed, for a particular application, it might be important to consider many different parameter choices in our generalized quality function.  We do not worry about such details here, as our goal has been to present a framework that allows one to study the community structure of multislice networks, but it is nevertheless important to mention it for further consideration.  We additionally note that optimizing the standard modularity quality function is known to be an NP-complete problem (\emph{S4}), and the cautionary observations regarding modularity optimization \cite{resolution,good2009} naturally also apply to our more general multislice framework.


\subsection*{Examples}

We conclude by providing additional details for the three examples discussed in the main text.

\subsection*{Community Detection Across Multiple Scales}

We performed simultaneous community detection across multiple resolutions (scales) in the well-known Zachary Karate Club benchmark network, which encodes the friendships between 34 members of a karate club at a U.S. university in the 1970s \cite{karate}. Keeping the same 34-node unweighted adjacency matrix across slices (so that $A_{ijs}=A_{ij}$ for all $s$), the resolution associated with each slice is dictated by a value from a specified sequence of $\gamma_s$ parameters, which we chose to be the 16 values $\gamma_s=\{0.25, 0.5, 0.75, \ldots, 4\}$. In Fig.~2, we depict the community assignments that we obtained when the individual nodes are coupled with strengths $\omega=\{0, 0.1, 1\}$ between each neighboring pair of the 16 ordered slices. For each $\omega$, we took the higher quality partition from that given by a spectral method plus Kernighan-Lin (KL) node-swapping steps \cite{richardson2009,newmodlong} and a generalization of the Louvain algorithm (\emph{S5}) plus KL steps.
We note that, despite this approach, the depicted $\omega=1$ partition can be clearly improved by leveraging the definition of the inter-slice coupling; specifically, the communities of nodes 30-34 (in the renumbering in Fig.~2) at different resolutions can be merged to improve the total quality of the multislice partition. Future algorithmic improvements could explicitly identify similar situations where merging or breaking communities across slices might improve the overall quality.

When $\omega=0$, the optimal partition obtained corresponds to the union of the independent partitions of each separate resolution parameter. As $\omega$ is increased, the coupling between neighboring slices encourages the partition to include communities that straddle multiple slices in the hierarchy of scales. The mathematical limit of arbitrarily large $\omega$ requires that, eventually, the communities span the full range of the considered resolutions. Because only the resolution parameters differed from one slice to the next in this multiple-resolution example, the limit of infinitely large inter-slice coupling here corresponded to single-resolution community detection at the average of the selected $\gamma_s$ values, $\langle\gamma_s\rangle \approx 2.125$.  Even at the smallest value of the resolution parameter that we used ($\gamma = 0.25$), we already observed a split into two communities when $\omega > 0$ (recalling that the actual club fractured into two groups).  We simultaneously obtained all of the other network scales, such as the partitioning of the Karate Club into four communities at the default resolution of NG modularity \cite{fortunato2009,richardson2009}. We also identified nodes that have an especially strong tendency to break off from larger communities (e.g., nodes 24--29 in Fig.~2).

This example illustrated that multislice community detection makes it possible to systematically track the development of multiple network scales simultaneously.


\subsection*{Community Detection in Time-Dependent Networks}

We considered roll call voting in the United States Senate across time. The Senate is one of the two chambers of the legislative branch (collectively called the Congress) of the U.S. federal government.  It currently consists of 100 Senators (two from each state) who serve staggered six-year terms such that approximately one-third of the Senate is elected every two years. The data we studied is from the 1st--110th Congresses, covering the years 1789--2008 and including 1884 individual Senators.\footnote{At least five Senators in the data [available at \url{voteview.com}\cite{voteview}] are each assigned two different identification numbers, corresponding to different periods of their careers.  We take the data as provided, counting such Senators twice, and merely remark that politically-minded studies should include such considerations.}  With each slice (i.e., within each two-year Congress), we defined a weighted connection between each pair of Senators in terms of a similarity between the votes they cast during that Congress\cite{waugh2009}. We then demonstrated that one can gain additional understanding of this network, and the underlying political processes, by applying multislice community detection to the collection of these 110 network slices taken as a whole.  In this multislice network, we coupled each individual Senator to him/herself when appearing in consecutive Congresses.  If a Senator from Congress $s$ did not serve in Congress $s+1$, then we did not introduce inter-slice coupling between slices $s$ and $s+1$ for this individual. With this formulation, link strengths and nodes (Senators) both changed from one slice to another.

Multislice community detection uncovered details about the
individual and group voting dynamics over time that are
simply not captured by the union of the 110 independent partitions of
the individual Congresses. Again using a generalization of the Louvain
algorithm plus KL steps, and using
inter-slice coupling $\omega=0.5$,
we obtained the partition depicted in Fig.~3
of the 1884 unique U.S.~Senators in
each Congress in which they voted into 9 communities. This community structure highlights several historical turning points in U.S. politics. For instance, the Congresses in which three communities appeared
simultaneously are each historically significant: The 4th and 5th
Congresses were the first with political parties; the 10th and 11th
Congresses occurred during the political drama of former Vice President Aaron
Burr's indictment for treason; the 14th and 15th Congresses witnessed
the beginning of changing group structures in the
Democratic-Republican party\cite{waugh2009} amidst the dying
Federalist party; the 31st Congress included the Compromise of 1850;
the 37th Congress occurred during the beginning of the American Civil
War; the 73rd and 74th Congresses followed the landslide 1932 election
amidst the Great Depression; and the 85th--88th Congresses brought the
major American civil rights acts, including the Congressional fights
over the Civil Rights Acts of 1957, 1960, and 1964
(observe that all 44 Democratic Senators in the community
colored green during this time period came from Southern states).  A
more complete political study using multislice community detection, which includes systematically examining the community structure as the inter-slice coupling strength $\omega$ is varied, would enable one to investigate such observations in extensive detail.


\subsection*{Community Detection in Multiplex Networks}

We applied multislice community detection to a multiplex network of 1640 college students at an anonymous, northeastern American university \cite{T3}. We included the symmetrized connections from the first wave of this data (covering the first year of university attendance) representing (1) Facebook friendships; (2) picture friendships, in which a student posted and tagged a photograph of another online; (3) roommates, in which two students shared a first-year dormitory room, creating clusters of 1--6 students; and (4) ``housing group'' preferences identified by the students.
Because the different tie types are categorical, the natural inter-slice couplings connect an individual corresponding to one type of connection to him/herself in each of the other 3 types of networks.  This type of inter-slice coupling thus has a different nature from the inter-slice couplings above that connected only neighboring (ordered) network slices.

In Table~\ref{table:T3}, we provide a summary of the basic results that we obtained by varying the inter-slice coupling strength $\omega$.  We tabulated the total number of communities and the percentages of individuals assigned to 1, 2, 3, or 4 communities in the multislice network across the four types of connections. Again, $\omega=0$ yielded separate communities for each slice, as expected, with each individual placed into four separate communities.
As $\omega$ was increased, communities merged across slices---most predominantly where the patterns of connection were relatively similar between two slices. This reduced the total number of communities and resulted in individuals with fewer distinct community assignments across their 4 appearances in the different network slices. For $\omega \in [0.2,0.5]$, a significant majority of the individuals were assigned to only 1 or 2 communities, indicating that their social networks maintain group-level similarities across the four types of connections. Another significant set of students were grouped into 3 different communities, and a small minority maintained 4 separate assignments, suggesting stark differences in their positions in the 4 single-category network slices. Finally, for $\omega = 1$, the inter-slice coupling was sufficiently strong that it forced all 4 multislice nodes corresponding to an individual student to be assigned to the same community. Further investigation of such different community assignments across slices could be used to more clearly compare and contrast the roles of individuals in each network slice and in the complete multislice network.
Additionally, a multislice approach might provide a novel mechanism for dealing with the problem of overlapping community assignments, as the hard partitioning of each node (located in a single slice) in the multislice network allows an individual to be placed into different communities in their appearances in the different slices.  Indeed, multiplexity is itself a strong motivation for developing methods that allow communities to overlap (\emph{2,3,S6}).




\section*{Supplementary References}

\begin{itemize}
\item[S1.] J.~Moody, DuPRI working paper PWP-DUKE-2009-009, \url{http://papers.ccpr.ucla.edu/abstract.php?preprint=722} (2010).
\item[S2.] J.~C. Delvenne, S.~N. Yaliraki, M.~Barahona, {\it arXiv:0812.1811\/}  (2008).
\item[S3.] Y.~Kim, S.-W. Son, H.~Jeong, {\it Physical Review E\/} {\bf 81}, 016103 (2010).
\item[S4.] U.~Brandes, D.~Delling, M.~Gaertler, R.~Goerke, M.~Hoefer, Z.~Nikoloski, D.~Wagner, {\it IEEE Transactions on Knowledge and Data Engineering\/} {\bf 20}, 172 (2008).
\item[S5.] V.~D. Blondel, J.-L. Guillaume, R.~Lambiotte, E.~Lefebvre, {\it Journal of Statistical Mechanics: Theory and Experiment\/} {\bf 2008}, P10008 (2008).
\item[S6.] G.~Palla, I.~Der\'{e}nyi, I.~Farkas, T.~Vicsek, {\it Nature\/} {\bf 435}, 814 (2005).
\end{itemize}

\end{document}